# The Time Programmable Frequency Comb: Generation and Application to Quantum-Limited Dual-Comb Ranging

Emily D. Caldwell[1], Laura C. Sinclair[1], Nathan R. Newbury[1], Jean-Daniel Deschenes[2]

[1]National Institute of Standards and Technology (NIST), Boulder, CO, USA

[2]Octosig Consulting, Quebec City, Canada

**The classic self-referenced frequency comb acts as an unrivaled ruler for precision optical metrology in both time and frequency[1,2]. Two decades after its invention, the frequency comb is now used in numerous active sensing applications[3–5]. Many of these applications, however, are limited by the tradeoffs inherent in the rigidity of the comb output and operate far from quantum-limited sensitivity. Here we demonstrate an agile programmable frequency comb where the pulse time and phase are digitally controlled with ±2 attosecond accuracy. This agility enables quantum-limited sensitivity in sensing applications since the programmable comb can be configured to coherently track weak returning pulse trains at the shot-noise limit. To highlight its capabilities, we use this programmable comb in a ranging system, reducing the detection threshold by ~5,000-fold to enable nearly quantum-limited ranging at mean pulse photon number of 1/77 while retaining the full accuracy and precision of a rigid frequency comb. Beyond ranging and imaging[6–12], applications in time/frequency metrology[1,2,5,13–18], comb-based spectroscopy[19–27], pump-probe experiments[28], and compressive sensing[29,30] should benefit from coherent control of the comb-pulse time and phase.**

As applications of frequency combs have expanded, their uses have extended beyond functioning simply as a reference ruler[3–5]. For example, many experiments combine two or more frequency combs for active sensing including precision ranging and imaging[6–12], linear and non-linear spectroscopy[19–27], and time transfer[13–18,31,32]. In these applications, the multiple fixed combs serve as differential rulers by phase-locking them to have a vernier-like offset between their frequency comb lines, or their pulses in time. These applications do exploit the accuracy and precision of frequency combs; however, despite the use of heterodyne detection, they operate nowhere near the quantum (or shot noise) limit because of effective dead time due to sensing the incoming signal-comb light via a comb with a deliberately mismatched repetition frequency. Consequently, there are strong tradeoffs in measurement speed, sensitivity and resolution[19,33,34]. In some dual-comb ranging and spectroscopy demonstrations, these penalties have been partially addressed by incoherent modulation of the comb[35–38] but not eliminated.

Here, we overlay a self-referenced optical frequency comb with synchronous digital electronics for real-time coherent control of the comb's pulse train output. We manipulate the frequency comb's two phase locks to dynamically control and track, the time and phase of the frequency comb's output pulses at will. The temporal placement of the comb pulses is set with ±2 attoseconds accuracy and with a range limited only by slew rate considerations. This time programmable frequency comb (TPFC) goes beyond the "mechanical gear box" analogy often applied to optically self-referenced combs[5], replacing it with a digitally controllable, agile, coherent optical pulse source. The agility of the TPFC enables many more measurement modalities than a rigid frequency comb. In sensing applications, the TPFC can enable quantum-limited detection with the full accuracy and precision of the frequency comb, avoiding the penalties discussed previously. To achieve these combined advantages, the TPFC is configured as a tracking optical oscillator in time and phase so that it effectively locks onto an incoming weak signal pulse train for coherent signal integration.

As an immediate example, we incorporate the TPFC into a dual-comb ranging system. The result is quantum-limited sensing that sacrifices none of the exquisite accuracy and precision of frequency-comb measurements. Here, we show a precision floor of 0.7 nm (4.8 attoseconds in time-of-flight) in ranging, which exceeds previous conventional dual-comb ranging demonstrations[6–8,39–41]. In addition, the tracking dual-comb ranging detects a weak reflected signal-comb pulse-train with a mean photon number per pulse of only 1/77 at a sensitivity within a factor of two of the quantum limit. Quantum-limited detection of signals at even lower mean photon per pulse numbers are possible by reducing the measurement bandwidth, depending on the application. In contrast, conventional dual-comb ranging would require a return signal 37 dB or 5000x stronger to reach the same level of performance.

The uses of the TPFC go well beyond acting as a tracking optical oscillator. It should enable many more time-based measurement schemes than the conventional vernier approaches using fixed frequency combs. For example, in multi-comb sensing, the relative time offset between the frequency combs can be adjusted so as to effectively mimic a higher repetition rate system while retaining the benefits of a lower repetition rate system, *e.g.* higher pulse energy and tight stabilization. Arbitrary patterns can enable future compressive sampling[30]. In time/frequency metrology, the comb can provide accurately adjustable timing signals, modulation capabilities for noise suppression, and optically-based time interval standards[42]. Multiple TPFCs could be used for pump-probe experiments with digital control of pulse spacing replacing delay lines or chirp-induced delays[28].

In this article, we first describe the TPFC and its capabilities generally. We then explore a specific application by integrating the TPFC into a dual-comb ranging system. Finally, we discuss the potential benefits of a TPFC in comb-based sensing more generally, including in LIDAR, spectroscopy, and time transfer.

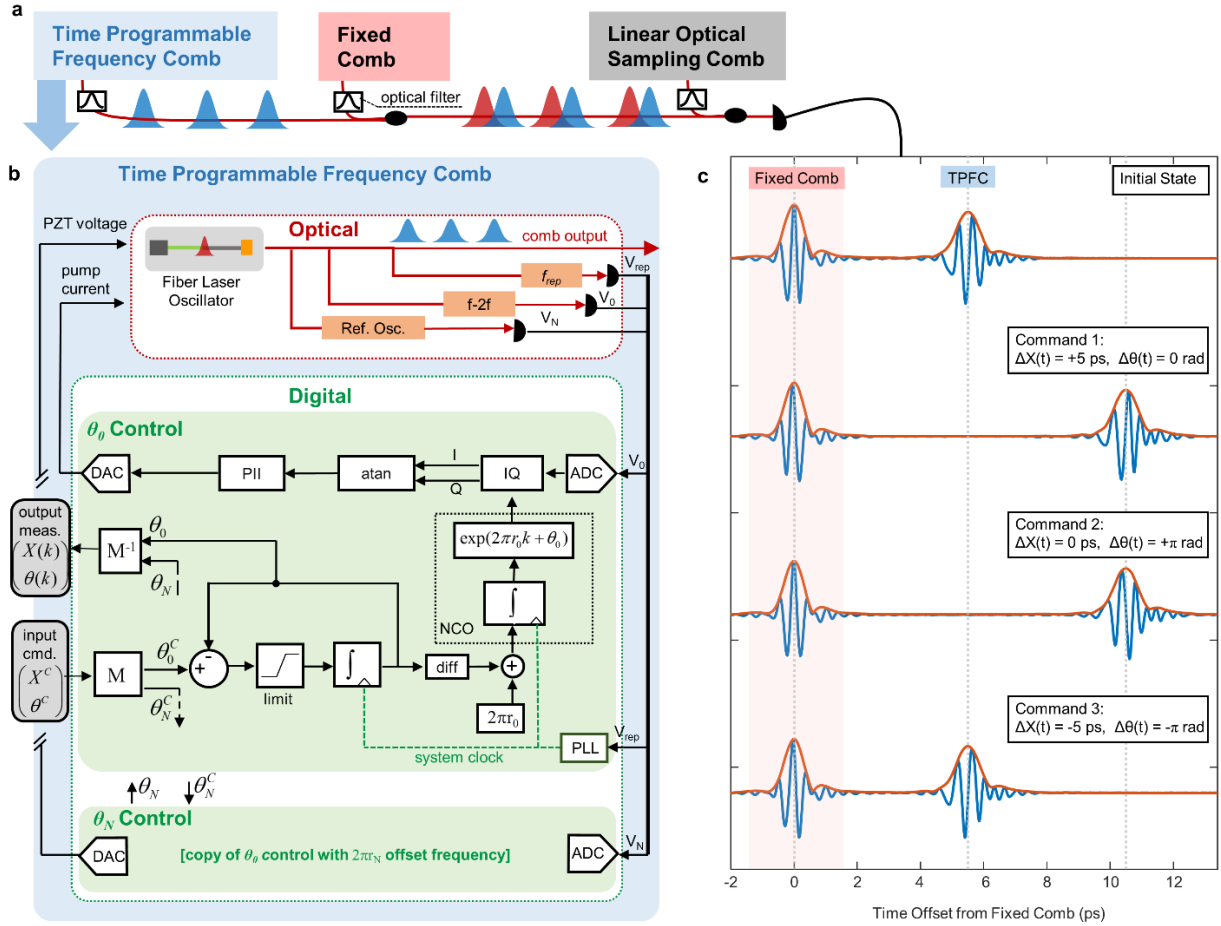

**Figure 1: A time programmable frequency comb (TPFC).** (a) The TPFC output is measured with respect to a second fixed frequency comb through linear optical sampling against a third frequency comb with an offset repetition frequency. All frequency combs are fiber-based operating at $f_{rep}$ ~200 MHz with a resulting 5-ns pulse spacing. All pulses are spectrally filtered to a Gaussian 10.1-nm wide shape, corresponding to 355 fs pulse duration (See Methods). (b) Schematic of the TPFC made from a self-referenced Er:fiber frequency comb controlled with digital electronics clocked off the detected comb repetition rate signal ($V_{rep}$). The digital section receives the carrier-envelope offset signal ($V_0$), the optical beat signal ($V_N$), and the comb pulse timing and phase commands, $X^C$ and $\theta^C$, which are combined to give the control phases $\theta_0^C$ and $\theta_N^C$ through the (trivial) matrix M. These are passed to their respective digital control loop (see Methods). The control efforts for $\theta_0^C$ and $\theta_N^C$ adjust the phase-locked loops (PLLs) controlling the comb's two degrees of freedom. The system tracks the actual phases, $\theta_0$ and $\theta_N$ as fixed point numbers, which are combined to give the actual pulse timing and phase, $X(k)$ and $\theta(k)$, for every comb pulse number $k$. IQ: in-phase/quadrature demodulator, PII: proportional-integral-integral controller, NCO: numerically controlled oscillator, $r_0$ and $r_N$: offset frequencies of the $\theta_0$ and $\theta_N$ phase locks in units of $f_{rep}$ (see Methods). (c) Linear optically sampled signal of the fixed comb (at $X = 0$) and the TPFC at the given $(X, \theta)$ values. Here, time is scaled such that $f_{rep}$ = 200 MHz.

## RESULTS

**Generation of a Time Programmable Frequency Comb**

The TPFC requires two parts: an *optically* self-referenced frequency comb and the electronics to track and control the time and phase of the comb pulses. (See Methods Eqn. 3 for a definition of the time and phase of the comb pulses.) While the electronic system need not be exclusively digital, it does need to track the programmed comb time and phase at the attosecond level over long (hours to weeks) durations. Here, we use a fixed-point number whose least significant bit corresponds to < 1 attosecond shift in time. When combined with an integer pulse number in an 80-bit number, the pulse timing can be specified with zero loss of accuracy for over 1 week at 1-as precision, thereby providing well beyond $10^{19}$-level control of the comb timing commensurate with next-generation optical clocks. As for the comb, any self-referenced comb could be converted into a TPFC; here, we generate a TPFC using a fiber-based comb.

Figures 1 and 2 describe the TPFC and its output characterization. In a self-referenced comb, phase-locked loops (PLLs) stabilize the frequency of the $N^{\text{th}}$ comb tooth, $f_N$, with respect to a CW reference laser, and the frequency of the $0^{\text{th}}$ comb tooth, $f_0$ (the carrier-envelope offset frequency). The PLL locks both frequencies to a known fraction of $f_{rep}$, which is self-referentially defined as $f_{rep} \equiv (f_N - f_0)/N$ [1,2,4,5]. These PLLs also set the phases of the $N^{\text{th}}$ and $0^{\text{th}}$ comb tooth frequencies, $\theta_N$ and $\theta_0$, to arbitrary but fixed values. Here, we manipulate these phases to control both the comb-pulse phase, $\theta$, and the comb-pulse time offset which is given by $X = (\theta_0 - \theta_N)/(2\pi N f_{rep})$ in direct analogy to $f_{rep}$'s definition above. The digital control exploits the optical frequency division of $N$ inherent for optically self-referenced combs since a single $2\pi$ shift in the phase of either PLL leads to a time shift ~ 5 femtoseconds. The TPFC outputs both a train of optical pulses and the corresponding synchronous digital values (Fig. 1b).

The TPFC is both agile and accurate (Figure 1c and Figure 2): at any point the output time of a comb pulse can be adjusted arbitrarily. Yet at all points, we know exactly, to fractions of an optical cycle, by how much the output time (and phase) has been shifted. For rapid changes in the TPFC output, the settling time of the PLLs can be taken into account either via modeling or by including the digital phase error signal from the two PLLs. It is the exactness of the performed step relative to the commanded step (Fig. 2b) and the ability to control the steps in real time that stands in contrast to earlier work. Shown in Fig. 2b, the accuracy of the timing control, $X$, with respect to the underlying CW reference laser is 0.77 ± 2.05 attoseconds. The TPFC enables easy determination of the mode number $N$, which is always required in optical frequency metrology, by applying a shift $\Delta\theta_0 = 2\pi N f_{rep}$, which will lead to a time shift of exactly $\Delta X = f_{rep}^{-1}$ for the correct $N$. Any integer error in $N$ appears as a 5-fs offset in time, which is easily resolved.

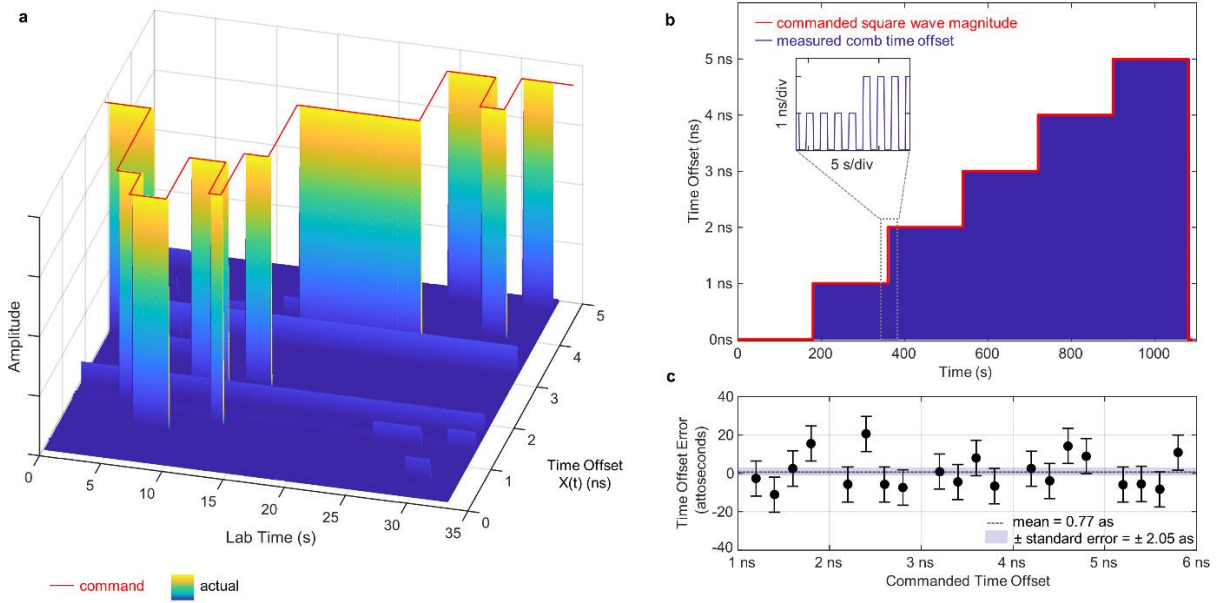

**Figure 2: Measurements of the TPFC pulse time, $X(t)$, when stepping $\theta_0(t)$, using the sampling setup in Fig 1a.** (a) Actual time offset between the TPFC and fixed comb for an arbitrary $X(t)$ step pattern (surface plot) compared to the commanded time (solid red line). (b) Staircase modulation of the comb pulse timing to verify the accuracy. We apply a 1-Hz square wave modulation to $X(t)$ with an amplitude that is stepped by 1-ns every 3 minutes. The 1-Hz modulation removes measurement noise from drifts in the fiber optic paths in Fig. 1a. The TPFC position is then measured by the linear optical sampling (LOS) frequency comb at a 6 kHz update rate (blue trace) and compared to the commanded time step (red line). (c) The average difference between the actual and commanded pulse time for both the data in (b) and additional data runs. The uncertainty bars are based on the LOS measurement and residual comb timing jitter. The average difference is 0.77 attoseconds ±2.05 attoseconds (standard error). There is no observed reduction in accuracy or precision associated with the step scan despite moving the TPFC over the full 5 ns ambiguity range. The pulses can be successfully commanded to shift in time by more than 5 ns, although then an additional pulse slips in between. For these data, the maximum slew rate was 40 ns/s.

**Example Application: Dual-Comb Ranging with a TPFC**

To demonstrate the advantages of the TPFC in dual-comb sensing, we consider ranging[6–8]. In dual-comb ranging, pulses with bandwidth $\tau_p^{-1}$ from a comb are reflected off an object, and their time-of-flight is detected by coherently heterodyning them against a second comb. This time-of-flight has a resolution of $\Delta R = c\tau_p/2$ and a non-ambiguity range $R_{NA} = c/(2 f_{rep})$, associated with "which pulse" is detected. (This ambiguity can be removed by changing $f_{rep}$ and repeating the measurement[7]). The accuracy is set by the comb's reference oscillator or knowledge of the index of refraction. The precision is, at best, equal to the resolution divided by the SNR:

$$\sigma_R = \frac{C}{2\ln(2)} \frac{\Delta R}{SNR_S} \tag{1}$$

where the $(2\ln(2))^{-1}$ factor arises from the assumption of Gaussian pulses (see Methods). The shot-noise limited signal-to-noise ratio, $SNR_S = \sqrt{\eta n_s}$ where $\eta$ is the detector quantum efficiency and $n_s = P_{rec}T/(h\nu)$ is the number of signal photons for a received power $P_{rec}$ and integration time, $T$. The constant $C$ quantifies how far the precision is from the quantum limit. It can be related to the power penalty as $P_p = C^2$. An optimal quantum-limited ranging system operates at $C = P_p = 1$.

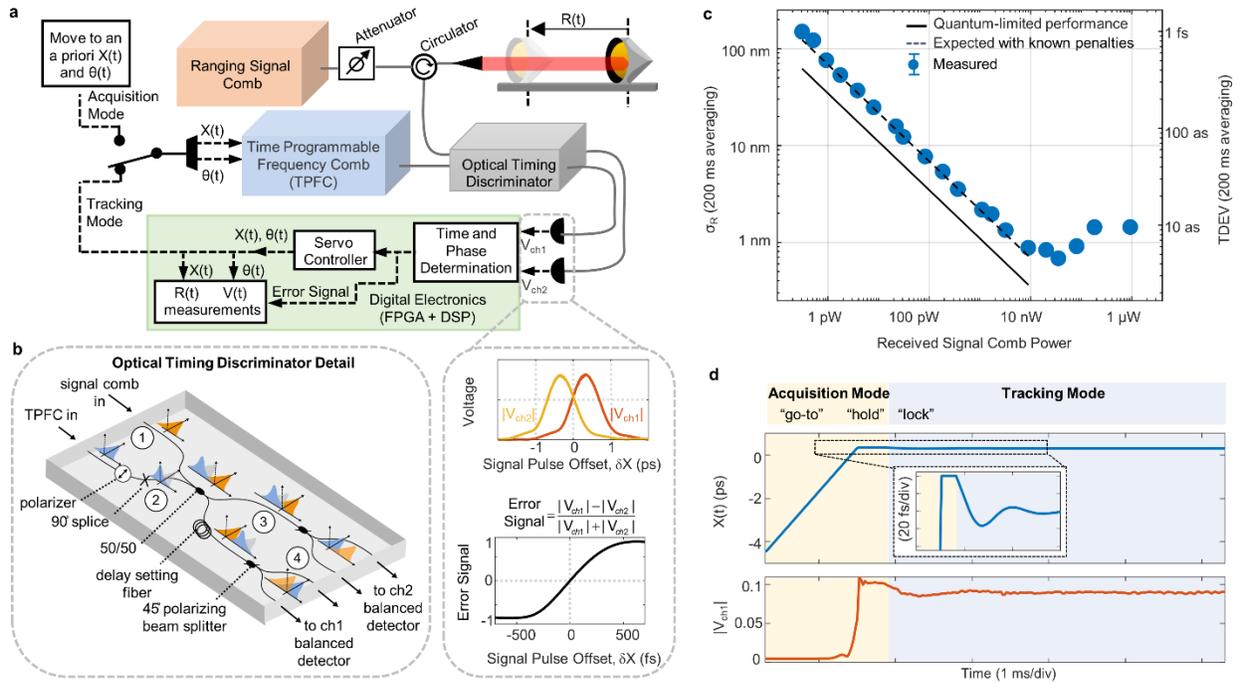

**Figure 3: Dual-comb ranging with a time programmable frequency comb.** (a) System diagram. The TPFC can be run in two modes: Acquisition Mode where the TPFC is moved to an *a priori* X(t) and θ(t), and Tracking Mode where feedback maintains temporal overlap between the TPFC and returning signal-comb pulses. By combining the feedback control effort and error signal for X(t) we measure the range, R(t). The velocity, V(t), is derived from dθ(t)/dt. (b) The timing discriminator is a dual Mach-Zehnder interferometer constructed with polarization-maintaining fiber optics in which (1) both comb pulses enter with the same polarization and (2) the TPFC pulse is rotated to the fast-axis. Then (3) the pulses are mixed, a delay between pulses is added to one arm and (4) the pulses are projected back into the same polarization for balanced heterodyne detection. The output signals for the two timing discriminator channels, $V_{ch1}$ and $V_{ch2}$, are combined to generate a power-insensitive error signal fed back to the tracking controller. (c) Range precision (deviation) $\sigma_R$ (left axis) and corresponding time deviation (TDEV, right axis) at 200-ms averaging time versus the total received signal-comb power at the balanced photodetectors. The measured precision follows the quantum limit (Eq. (1)) from 0.33 ± 0.03 pW to 10 ± 1.0 nW with a penalty of *C* = 2.16, reaching a systematic noise floor below 1 nm (7 as), which floor is 2-10x below previous dual-comb ranging experiments[6–8,39–41]. (See also Methods and Extended Data Figure 1.) (d) Example handover between Acquisition and Tracking modes, as described in the text. Here the TPFC was commanded to move from X(t) = -5 ps to X(t) = 0 ps where the signal pulse is located, and then lock onto the signal pulse.

Conventional dual-comb ranging operates far from this optimum unity power penalty (later Fig. 5) and with significant tradeoffs. In these systems, the second comb's repetition rate is offset by $\Delta f_r$ and it serves as a linear sampling comb. It repeatedly scans the entire non-ambiguity range, $R_{NA}$, at a measurement rate $T^{-1} = \Delta f_r \leq f_{rep} \Delta R / (4 R_{NA})$ (Extended Data Table 1). The inherent tradeoffs in $T$, $R_{NA}$, and $\Delta R$ have led to dual-comb ranging implementations using very different frequency combs and covering three orders of magnitude in $T$ and $\Delta R$ but all facing these strong constraints [6-8,39-41]. Moreover, in all cases, the power penalty $P_p \approx \Delta R / R_{NA} = f_r \tau_p$ is severe, ranging from 14 dB to 38 dB[7,39], because of the repeated scanning of the entire non-ambiguity range.

Here, as shown in Figure 3, we replace the second sampling frequency comb by the TPFC to overcome these tradeoffs and stiff power penalty. This system runs in two modes: acquisition mode and tracking mode, both of which differ from conventional dual comb ranging. In both modes, the relative time and phase between the TPFC pulses and the strongly attenuated signal-comb pulses and are detected by a power-insensitive, coherent timing discriminator (Fig. 3b), followed by a 26-kHz bandpass filter for coherent averaging. (The 26 kHz bandwidth was chosen to be well above typical ~kHz mechanical vibrations, but lower bandwidths are also possible.) The timing discriminator effectively shapes the TPFC pulse to optimize detection of the time of the incoming signal-comb pulses,[43,44] independent of their energy.

In acquisition mode, X(t) is scanned until the tracking comb's timing matches the input pulse train. While it is possible to scan the entire non-ambiguity range, we can also make use of *a priori* information to scan the TPFC's $X(t)$ linearly over a tunable range that can be much less than the non-ambiguity range. The information could be provided from external sources or from a Kalman filter if previous range/Doppler measurements are available. Once the system acquires the appropriate reflection, it switches to tracking mode (Fig. 3d). Tracking mode implements a pulse-timing lock and a carrier-frequency lock based on the timing discriminator outputs. The former keeps the signal and tracking comb pulse trains overlapped in time at the detector by adjusting $X(t)$. The latter adjusts for Doppler shifts in the return signal by applying feedback to the direct digital synthesizer used in demodulating the timing discriminator signals. The combination of the control and error signals from the time and frequency locks in turn yield the range and Doppler velocity of the target object.

In tracking mode, the ranging precision nearly reaches the quantum limit of Eq. (1) (Fig. 3c). This nearly quantum-limited precision ranging is demonstrated at a rapid 26-kHz measurement rate with as little as 0.33 ± 0.03 pW of return power ($SNR_S$ = 9.5), which corresponds to only 0.013 mean photons per pulse. There is a slight penalty of $C$ = 2.16x due primarily to differential dispersion between the comb pulses (see Methods). With additional optimization, $C$ could be reduced to 1, and with squeezing, to <1 per Ref. 43. With these same 200-MHz combs, a conventional dual-comb ranging would suffer from a power penalty $P_p$ = 37 dB ($C$ = 71). Finally, momentary loss of signal is not an issue. If brief enough, the object does not move by more than ~ $\pm 2 \Delta R$ = ~±100 μm and the acquisition resumes. If longer, the system can transition to acquisition mode using previous data to limit the scan, as discussed above.

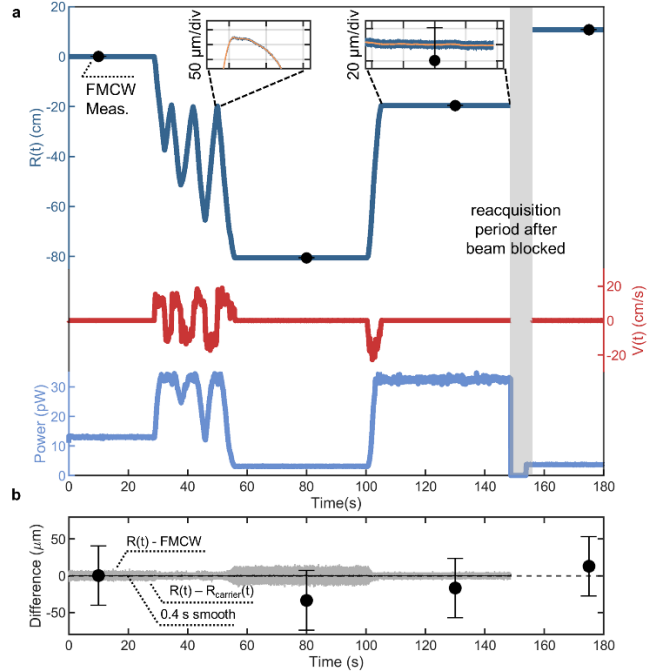

**Figure 4: Ranging and velocity data for a moving retroreflector.** (a) The range (top, left axis, dark blue trace) is measured from the summed control and error signals for X(t) in tracking mode at a 26 kHz rate. The velocity (right, middle axis, red trace) is based on the measured Doppler shift between the signal and TPFC combs calculated from the θ(t). At 150 seconds, the beam was blocked and the target moved, triggering a re-acquisition. Before and after re-acquisition the measured range agrees with a commercial FMCW ranging system (black circles), to within the FMCW uncertainty driven by target vibrations (see Methods). The relative range also agrees well with the unwrapped carrier phase (yellow trace in the two insets), after applying an overall offset. (b) Difference between the dual-comb range and FMCW range (black circles) with standard deviation error bars and difference between the absolute range from $X(t)$ and the unwrapped carrier phase (grey and black traces) for the initial period of continuous signal. The range deviation of this latter difference reaches 10 nm and 5 nm at 10-s averaging for the time periods with received powers of $3.2 \pm 0.3$ and $32 \pm 3.0$ pW, respectively. (See Extended Data Figure 1.)

Figure 4 shows range data taken while arbitrarily moving the rail-mounted retroreflector. The dual comb system tracks the retroreflector as it reverses direction at velocities up to 20 cm/s. The signal is blocked at 150 s and the retroreflector moved, after which the absolute range was reacquired by scanning over a ±37.5 cm window. The frequency shift of the returning reflection yields the velocity. To validate absolute ranging, we compare to a commercial frequency-modulated continuous-wave (FMCW) system at a few static rail positions after calibrating out differential range offsets. The two agree to within the FMCW measurement uncertainty of ±40 μm whose extent is due to target vibrations amplified by the FMCW's intrinsic range-doppler coupling (Extended Data Figure 2). Finally, the tracking dual-comb system also outputs the phase, $\theta(t)$, whose derivative yielded the velocity above. However, the relative phase, $\theta(t) - \theta(0)$ can also be unwrapped to provide relative range during periods of continuous signal (Fig. 4, yellow trace) as in Ref. 7 and similar to CW interferometry (except avoiding systematic errors from spurious reflections). This unwrapped carrier phase agrees with the tracking range to a precision limited by the tracking range noise and follows Eq. (1), after accounting for a ~1.5x chirp-induced penalty in $C$ from the fiber optic path to reach the rail system.

**DISCUSSION**

A number of existing or potential applications should benefit from the abilities illustrated in Fig. 1-3, specifically to (1) set the time and phase of the comb's output pulses, (2) coherently scan the relative temporal spacing between two frequency combs over a specified limited range rather than the full inverse repetition rate, thereby mimicking a higher repetition-rate comb while avoiding limitations of lower pulse energy, and (3) operate as a precision optical tracking oscillator in time and frequency for shot-noise limited sensing. Below we discuss three different general application areas: LIDAR, time metrology, and spectroscopic sensing.

As already discussed, frequency combs have a natural connection to precision LIDAR. Figure 5 and Extended Data Table 1 together compare conventional dual-comb ranging[7,39–41], tracking dual-comb ranging and FMCW ranging[45], which is the standard approach to high-resolution optical ranging. For all three, the resolution is set by the optical bandwidth and the accuracy by the comb referencing or knowledge of the index of refraction of air. (Comb-assisted FMCW ranging can transfer frequency-comb accuracy to FMCW LIDAR[46].) Both tracking dual-comb ranging and FMCW ranging can reach the shot-noise limit and exploit the optical carrier phase. However, FMCW ranging's update rate is limited by the need to sweep the laser over the ~THz bandwidth. Moreover, it is also much more sensitive to vibration because of its diagonal range-Doppler ambiguity function, which limits its precision for vibrating targets as shown in Figure 4. Because dual-comb ranging operates directly in the time-domain, it could provide range-resolved vibrometry in a cluttered environment or image through turbulent media. Moreover, tracking dual-comb ranging could provide quantum-limited surface imaging since it can operate at very high loss and should be robust to signal dropout from speckle. At 10-mW launch power, even a -100 dB reflection from a diffusely scattering target will still provide sufficient 1-pW return power to enable a 26-kHz measurement rate. More generally, there are strong overlaps with conventional RF pulse-Doppler radar and the tracking comb could therefore have interesting applications to high-bandwidth synthetic aperture LIDAR[47].

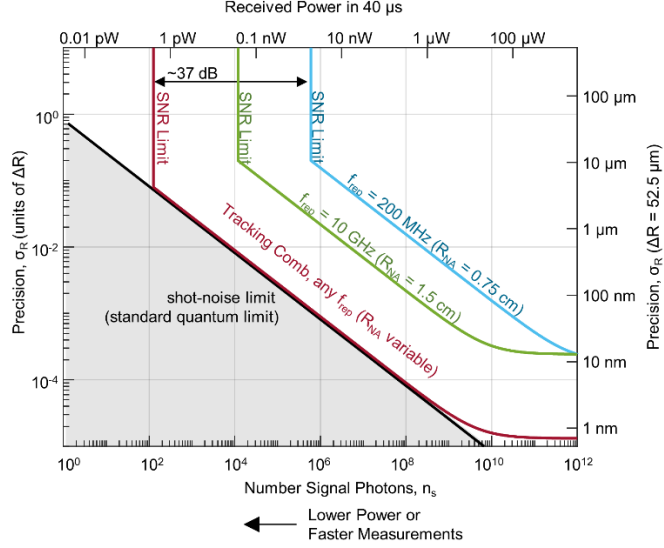

**Figure 5: Precision versus received signal photons,** $n_s = SNR_S^2/\eta$, **for conventional dual-comb ranging at** $f_{rep}$ **= 200 MHz (blue line) and** $f_{rep}$ **= 10 GHz (green line), compared to tracking dual-comb ranging (red line) and the standard quantum limit for heterodyne detection (black line). The left/bottom axes are in normalized units while the right/top axes are scaled to values similar to those used here. For all three ranging configurations, we require a minimum detection SNR of ~10, as indicated by the SNR limit. For conventional dual-comb ranging (blue and green lines), the scaling depends strongly on** $f_{rep}$ **and the precision is always much worse than the standard quantum limit. The tracking dual-comb ranging (red line) is independent of** $f_{rep}$ **and can reach the standard quantum limit, though here shown with a 10% penalty. We assume a system noise floor of 0.7 nm, taken from this work, for the tracking comb, and ~10 nm for the conventional dual-comb systems**[7,39]**.**

In time-frequency metrology, a TPFC phase-locked to an optical atomic clock provides an optical timescale with the ability to "adjust" its output time to synchronize with other signals. A TPFC could also enable calibration of time interval counters or ranging instruments[42]. The time-interval standard would follow Fig. 1c, where the TPFC allows precisely defined variable pulse spacing from nanoseconds to femtoseconds. This capability offers the prospect of a time interval standard that spans 6 orders of magnitude with attosecond precision and an accuracy directly tied to a secondary representation of the second. In secure optical communications, the programmable comb might enable quadrature pulse phase-position modulation if implemented with high-speed actuators. Finally, the TPFC has interesting applications to comb-based long-distance free-space time transfer [13–17,31,32] as it provides similar advantages as for dual-comb ranging[48].

The TPFC can also break tradeoffs that limit comb-based linear and non-linear spectroscopy. Relatively low repetition-rate frequency combs (100-MHz to 1-GHz) provide the high pulse energies needed for nonlinear spectroscopy or for spectrally broadening over the desired spectral band[49–51]. However, the spectral resolution set by these low repetition rates is often poorly matched to the application leading to significant "deadtime" or SNR reduction in multi-comb-based spectroscopy[19,21–26,50,51]. The TPFC can circumvent this problem by coherently scanning over a limited time offset between two or more frequency combs, as first demonstrated incoherently in the early dual-comb spectroscopic work of Schliesser *et al*.[35]. In this way, a low-repetition frequency TPFC can act as a frequency comb with an effectively much higher repetition rate, set by the inverse of the temporal scan window, at shot-noise limited sensitivity. Going further, the ability to jump the frequency comb pulse phase and timing could enable compressive sampling in dual-comb or multi-comb sensing applications with a concomitant increase in measurement rate. Recent modeling[30] and preliminary experiments[29] have highlighted the advantages of this dynamic control for dual-comb spectroscopy. Finally, in nonlinear spectroscopy, temporal control could enable time-ordered multi-photon excitation, following the comb-based spectroscopy of Rubidium[28] but with programmable control.

**CONCLUSION**

The time programmable frequency comb combines the precision and accuracy of a self-referenced frequency comb with flexibility in time and phase and 2-attosecond accuracy. Here, the TPFC is based on a fiber frequency comb, but in principle any self-referenced comb (or comb locked to widely separated optical oscillators) with control electronics capable of tracking and manipulating phase could act as a TPFC. Through a dual-comb ranging demonstration, we show the TPFC can operate as an optical tracking oscillator in time and frequency, yielding nearly quantum-noise limited ranging with 0.7 nm precision. Finally, dual-comb ranging is just one application and the TPFC has equal promise in relaxing tradeoffs with repetition frequency and improving SNR in other multi-comb sensing and metrology applications while retaining the hallmark accuracy of comb-based metrology.

**ACKNOWLEDGMENTS**
We acknowledge Jennifer Ellis, Tara Fortier, Kevin Cossel, William Swann, Benjamin Stuhl, and Brian Washburn for helpful discussions.

**AUTHOR CONTRIBUTIONS**
All authors contributed extensively to this work.

**MATERIALS AND CORRESPONDANCE**
Should be addressed to Laura Sinclair (laura.sinclair@nist.gov) or Nathan Newbury (nathan.newbury@nist.gov).

# METHODS

## I) Time Programmable Frequency Comb Control

The output of the frequency comb can be written as a function of time, $t$, in terms of a sum of pulses, labeled by integer $k$, with repetition frequency $f_{rep}$,

$$E(t) = e^{-i\theta} \sum_k E_k(t-X)$$
$$\text{where } E_k(t) = e^{-ik\theta_{ceo}} A(t - kf_{rep}^{-1}) \tag{2}$$

where $\theta$ and $X$ are the comb pulse phase and time offset (to be controlled later), and $\theta_{ceo}$ is the carrier-to-envelope phase advance per pulse (and hence appears as $k\theta_{ceo}$) and is not to be confused with $\theta$. Alternatively, through the Poisson sum formula, the comb can be written as

$$E(t) = f_{rep} e^{-i\theta} \sum_n \tilde{A}_n e^{-i2\pi f_n \cdot (t-X)}, \tag{3}$$

where $n$ is the index of the comb tooth with complex amplitude $\tilde{A}_n$ at frequency $f_n = nf_{rep} + f_0$ and the carrier-envelope offset frequency $f_0 \equiv (2\pi)^{-1} \theta_{ceo} f_{rep}$.

For a comb self-referenced to a reference CW laser, we stabilize the comb tooth frequencies, $f_0$ and $f_N$, where $N$ is the tooth nearest the reference CW laser at $f_{CW}$. Each of these frequencies is stabilized with respect to the repetition frequency, i.e. $f_0 = r_0 f_{rep}$ and $f_N = f_{CW} - r_N f_{rep}$, where both $r_0$ and $r_N$ are user-chosen rational fractions. Note that the repetition rate itself is determined by the two frequencies as $f_{rep} = (f_N - f_0)/N$. Equivalently, $f_{rep} = f_{CW}/(N + r_0 + r_N)$. This stabilization is done through two phase-locked loops that combined also set the overall phase, $\theta$, and time offset, $X$, to arbitrary but fixed values. If the phase and time offset were not fixed, variation in $X$ would lead to an effective variation in the repetition frequency $f_{rep}$, and variation in $\theta$ would similarly lead to a variation in $\theta_{ceo}$.

Here, we alter the lock points of the two phase-locked loops in order to coherently change the overall phase, $\theta$, and time offset, $X$. Consider the phase-lock of the tooth at $n=0$ (i.e. the carrier-envelope offset stabilization). The phases satisfy the equation,

$$[\theta + 2\pi f_0 t - 2\pi f_0 X] - r_0 [2\pi f_{rep} t - 2\pi f_{rep} X] - \theta_{0,cal} = 0 \tag{4}$$

where the first term in brackets is the phase of the $n=0$ term in (3) assuming an unchirped $A(t)$, the second term in brackets is the phase of the repetition signal, i.e. the digital clock signal for the digital phase-locked loop which directly follows the detected comb pulse arrivals times, and the final term is a calibration offset related to the difference in the total phase delays of the two signals up to the IQ demodulation (phase comparison). In the standard fixed comb control, feedback is applied to drive this phase difference to zero, as indicated by the right-hand side of the equation. With synchronous digital electronics, we can replace the right-hand side with a time-dependent control phase, $\theta_0^C(t)$ so that

$$\theta(t) = \theta_0^C(t) \tag{5}$$

since $f_0 \equiv r_0 f_{rep}$ and after dropping the calibration constant for simplicity. We apply the same arguments to the phase of the $n = N$ tooth, working from the stabilization of $f_N = f_{CW} - r_N f_{rep}$ to find,

$$\theta(t) - 2\pi(N + r_0 + r_N) f_{rep} X(t) = \theta_N^C(t). \tag{6}$$

A linear combination of Eqns. (5) and (6) (represented by the matrix $\mathbf{M}^{-1}$ in Fig. 1) gives us the simple relationships,

$$\theta(t) = \theta_0^C(t)$$
$$X(t) = \frac{\theta_0^C(t) - \theta_N^C(t)}{2\pi f_{CW}} \tag{7}$$

These equations relate changes in the time-varying control phases to the comb pulse phase and timing. They hold within the feedback bandwidth of the phase-locked loop. (In practice, we can change the signal more rapidly provided we separately record the error signal quantifying the difference between the set point and actual values.) For the particular locking conditions chosen here of $r_N = -r_0$, these equations reduce to the ones given in the main text. To set the TPFC output, the inverse of Eqn. (7) generates the desired values of $\theta_0^C(t)$ and $\theta_N^C(t)$ which are then translated to the actuator controls. (See Fig. 1 and below for details on processing).

With appropriate choice of the control phases, we can control the comb's time offset and phase, or the phase of a particular tooth through Eq. (8). The phase change of the $n^{th}$ tooth is

$$\theta_n(t) = \theta(t) - 2\pi f_n X$$
$$= \left(1 - \frac{n + r_0}{N + r_0 + r_N}\right)\theta_0^C(t) + \frac{n + r_0}{N + r_0 + r_N}\theta_N^C(t) \tag{8}$$

For the tracking comb operation, we adjust $\theta_0^C$ and leave $\theta_N^C = 0$. This leads to a time shift of $\Delta X = \Delta \theta_0^C / (2\pi N f_{rep})$ while fixing the phase of the $N^{th}$ comb tooth to zero even while the timing of the TPFC comb pulses is shifting. As a result, the relative phase shift of the $N^{th}$ comb tooth between the TPFC and signal comb reflects the overall additional phase shift on the signal comb alone. Therefore, in "unwrapping" the phase of the carrier signal measured during the retroreflector motion, we must use an effective carrier frequency of $Nf_{rep}$.

## III) Physical setup information on the combs and the digital electronics system

### III.A Optical Hardware

The fiber-based frequency combs used in this work operate at a repetition frequency of about 200-MHz and are based on the design in Ref. 2. The actuator for the carrier-envelope-offset phase-lock is the oscillator pump power and the actuators for the lock of the $N^{th}$ comb tooth to the optical reference are two piezo-electric fiber stretchers which together adjust the oscillator cavity-length. All combs are housed within temperature-controlled enclosures. The output of all combs are filtered with a 10.1-nm wide Gaussian filter centered at 1560 nm, resulting in 5 mW of in-band power, which was often strongly attenuated for these experiments. The optical reference for all combs is a cavity-stabilized laser at 1535.04 nm. To minimize excess residual noise between the frequency combs, all non-common fiber lengths between the cavity-stabilized laser and the three combs are as short as possible and contained within the temperature-controlled housing as much as possible.

In order to produce the required combination of TPFC, fixed comb and LOS comb pulses, fiber-optic-based optical transceivers were constructed with polarization maintaining fiber optics (PM-1550) and fiber optic components. The fiber-optic based timing discriminator described in Figure 1 was constructed with a differential pathlength of 50.5 cm of fiber which corresponds to a 980-femtosecond delay between the lead and lag arms to match the full-width-half maximum of the partially chirped cross-correlation (or interferogram) signal between the tracking and signal frequency combs. Commercial balanced detectors are used throughout, including at the output of the timing discriminator, to allow for shot-noise limited heterodyne detection with ~1.1 dB power penalty.

For the data of Fig. 1c and 2, we combine a fixed comb and a TPFC, both phase-locked with the same values of $N$, $r_0$, and $r_N$ We then adjust the TPFC pulse time or phase, and subsequently probe the relative phase and time between the TPFC and the fixed comb via a third linear optical sampling (LOS) frequency comb. For the data of Figures 3 and 4, we do not use the third frequency comb. In all cases, the LOS comb is phase-locked to the same CW reference laser but with carrier frequency offset by 10 MHz. The heterodyne signal between the LOS and the other two combs is digitized at $f_{rep,LOS}$ yielding an interferogram (IGM) with $f_{rep,LOS}/\Delta f_{rep}$ points. For the data of Figure 1c and 2a, the LOS comb was offset in repetition frequency by ~200 Hz while for the data of Figure 2b (which required faster averaging), the LOS comb was offset in frequency by ~6 kHz, very near the Nyquist limit for linear optical sampling given the pulse width. For Figure 2b we additionally apply a matched filter prior to peak fitting. In these LOS sampled measurements, the overlap between the LOS combs and the TPFC is sufficiently short that the full timing jitter of the relative comb pulses is measured. For these combs, this timing jitter was about 3 fs. For the data of Figure 2b, multiple measurements allow us to average down this jitter. At these longer averaging times, differential out-of-loop fiber paths in the TPFC and LOS combs will lead to slow fs-level time wander. We cancel some of this wander by applying the square-wave 1-Hz modulation to the TPFC time offset.

In contrast to the LOS sampling method, the combination of the TPFC and timing discriminator provides continuous measurements at each comb pulse. Low-pass filtering rejects most of the technical timing jitter on the comb pulses that occurs at ~ 50 kHz (the PLL locking bandwidth). The data for Extended Data Figure 1 and Fig. 3c was acquired using this continuous timing comparison between the reflected signal frequency comb and TPFC. For these data, the signal-comb pulses were reflected off a fiber FC/PC connector end located just outside the aluminum box that housed the two combs in order to minimize temperature-induced path length changes or the even larger atmospheric fluctuations that occur over the air path to the retroreflector.

For the ranging data of Figure 4, the retroreflector on the rail was aligned with 3 passes so that the effective pathlength change and Doppler-shift of the comb pulses were a factor of 3 higher than that of the reflector itself. The signal-comb light was attenuated significantly prior to launch to reach the low picowatt power levels. The position of the retroreflector on the rail was simply adjusted by hand in an arbitrary pattern. The alignment of the output signal-comb light to the retroreflector was not perfect, leading to the power variations in the return signal observed in Fig. 4.

At each static location, a commercial FMCW ranging system took 25 range profiles over a 10 second acquisition period. The reported FMCW range in Figure 4 is the average of the peak values extracted by a 3-point cubic spline fit of the peak. The uncertainty was generated by performing the same fitting routine over 10 minutes of static range data and taking the standard deviation of the resultant ranges. For both the dual-comb system and FMCW system, the range was calculated using the group velocity calculated at the center frequency. The large uncertainty on the FMCW range values is due to the systematic coupling between range and velocity in this frequency-domain ranging system. For the FMCW ranging system used here with a 12.5 THz/s sweep rate, the systematic ranging error is 12 μm x $\delta f_{Doppler}$, where $\delta f_{Doppler}$ is the vibration-induced Doppler shift. While Doppler shifts slower than the sweep rate can be compensated through the use of up/down sweeps, vibration-induced Doppler shifts on timescales comparable with the sweep time cannot be similarly cancelled, leading to the uncertainty bars on the FMCW ranging data in Figure 4. The tracking dual-comb ranging does not suffer from this strong Doppler-induced systematic since the ambiguity function for a pulse does not have the large delay-Doppler coupling of a slow FMCW sweep.

III.B Digital Electronics System

The implementation of the digital control of the TPFC was accomplished using available hardware and consists of two field programmable gate arrays (FPGAs) and one Digital Signal Processor (DSP), although these FPGAs and DSP could be combined into a single platform as the processing load is not significantly larger than fixed comb control. The FPGAs are both clocked synchronously at the comb repetition rate and the DSP processes samples synchronously at exactly 5000x slower (~40 kHz). Since the DSP is programmed in C++, it provides a more flexible development environment than the FPGAs. The fixed and LOS frequency combs also each had a dedicated FPGA[52] for their phase-locking to the CW reference laser.

To control the TPFC output, Eqn. (7) generates the desired values of $\theta_0^C(t)$ and $\theta_N^C(t)$. (Note that these are "unwrapped" quantities and are not restricted to a 0 to $2\pi$ range.) In the implementation, all phase and frequencies are defined with respect to the clock cycles of the analog-to-digital convertors (ADC), FPGAs and DSP that are synchronous with the instantaneous comb repetition frequency. Within the FPGA, the values of $\theta_0^C(t)$ and $\theta_N^C(t)$ can be represented as large fixed-point values such that there is no quantization noise variation across the full range (such as would be inevitable with floating points). Here, we use a least significant bit (LSB) that corresponds to a comb timing step of below 1 attosecond. All values are then otherwise exact and set with sub-cycle accuracy, assuming no cycle slips in any clock signal, which is assured by the high-SNR signal from the photo-detected frequency comb pulse-train.

Figure 1b shows the effective implementation for the phase trajectories for $\theta_0(t)$ and $\theta_N(t)$. The implementation is done via the numerical integration of a limited difference between desired and actual trajectories. This allows enforcing rate limiting to respect the limits of the physical system, for example the maximum rate that the system can change pulse timing via pump power or cavity length. The phase change is done by changing the effective values of $r_0$ and $r_N$ for a precisely known time period inside the FPGA. The commands for setting the trajectories can be sent via either a graphical user interface from a PC, via a serial port input from another real-time digital system for tighter control (e.g. the DSP). It could also be done via a sequencer running directly on the control FPGA, although that approach was not used for this manuscript. The rest of the control algorithm for the phase locks is a PII with a phase extraction front-end, as in Ref. 2.

For large timing shifts, the system will also be limited by the maximum slew rate. Note that a linear slew of the phase corresponds to a frequency shift in the relevant beat frequency. The slew rate can then be limited by two factors: a) how far the beat frequencies for either $f_0$ or the optical beat can be moved away from their nominal lock points while tracking the phase, and b) how fast the actuators can implement the required frequency changes on the physical system. For a), the limitation comes from a combination of SNR and sophistication of the demodulation scheme. Indeed, with unlimited SNR, the beats can be tracked over the full Nyquist window of 0 to $\pm f_{rep}/2$. If the SNR is limited, one could implement a tracking filter with feedforward to track dynamic frequency changes, although this was not implemented here. In the current system, a maximum slew rate of 40 ns/s was implemented although this rate was not optimized and was chosen conservatively to avoid cycle slips even with limited SNR. This 40 ns/s slew rate was used for the data of Figure 1 and 2. For the ranging data of Figure 3-4, a slower slew rate of 4 ns/s was chosen to match the temporal duration of the overlap between the TPFC and signal comb with the inverse of the 26-kHz measurement bandwidth.

With regard to the slew rate, the critical issue is that the system never loses track of the accumulated phase error: once that is accomplished, the feedback loop is linearized and can work to coherently follow the trajectory with respect to the optical reference despite limitations in the actuators. To check if the system can robustly program the comb's output without losing track of phase, e.g. without any cycle slips, the TPFC is programmed to repeatedly move back and forth between two target positions in time while comparing the results against the timing of a stationary comb pulse train. As noted in the text in the context of the removing the mode number ambiguity, even a single cycle slip will be evident as a measurable 5-fs shift in relative time offset.

For the dual-comb ranging demonstration of Figures 3 and 4, the system operates as nested phase-locked loops. The timing discriminator demodulation is implemented in an FPGA+DSP system with ±13 kHz of bandwidth centered around a demodulation frequency. The 26 kHz bandwidth in the demodulation limits the overall tracking bandwidth, but was chosen to demonstrate the high sensitivity by allowing for coherent integration over ~(26 kHz)$^{-1}$ = 38 μs to detect weak return signals. A broader bandwidth would sacrifice some sensitivity but allow for even faster tracking bandwidths. The output of the timing discriminator provides both a time and phase error signal. The phase error is sent to a PII implemented on the DSP, whose output is fed back to the demodulator, with the resulting frequency adjustment timestamped against the FPGA clock. (This timestamping allows the accurate unwrapping of the carrier phase to determine the relative range change, as show in Figure 4.) The time error signal is sent to a PII implemented on the DSP. This PII generates control trajectories for the tracking comb, which are sent via serial port to that comb's controller FPGA running the two PLLs for $\theta_0(t)$ and $\theta_N(t)$ at 50-kHz closed-loop bandwidth.

## **IV) Shot Noise Limit for Ranging**

In Fig. 3c, we show our ranging measurements are shot-noise limited for received powers <10 nW by comparing measured range deviation to the expected shot-noise limited range deviation. Here we derive that theoretical shot-noise limited range deviation.

We first assume the comb pulses have zero differential chirp and are Gaussian with a full-width half-maximum (FWHM) of $\tau_p$ in intensity. (The comb pulses pass through a Gaussian spectral filter before ranging.) Their cross-correlation then generates a Gaussian envelope, $V$, with a FWHM of $2\tau_p$ (because there is a factor of $\sqrt{2}$ from the cross-correlation and the E-fields are also $\sqrt{2}$ broader than the intensity of the comb pulses[1].) The detected heterodyne voltage signal has a carrier frequency set by the offset between the comb frequencies, but this carrier is removed in the demodulation to generate just the envelope signal. We assume the lock point is chosen at approximately the half-width at half-maximum (HWHM) point on this envelope. (The actual maximum slope is one-sigma from the peak, but this is a minor 6% correction at the cost of a reduced capture range.) In that case, the change in voltage for small changes in the arrival time of the signal pulse with respect to the tracking (LO) comb, $\delta X$, will be determined by the slope at the HWHM point, or

$$\frac{\delta V}{\delta X} = \ln(2) \frac{V_{\text{peak}}}{\tau_p} \quad [\text{V/s}] \tag{9}$$

where $V_{\text{peak}}$ is the maximum voltage measured when the signal and tracking comb pulses perfectly overlap. The inverse of this slope maps changes in $V$ to changes in $X$. Voltage fluctuations from shot noise with a root-mean squared (rms) value of $V_{\text{shot noise}}$ will show up as white timing jitter with an RMS value of

$$\sigma_{t,1} = \frac{\delta X}{\delta V} V_{shot\ noise}. \tag{10}$$

In the strong local oscillator case (strong tracking comb pulses), the shot noise for the demodulated output is

$$V_{shot\ noise} = eG\sqrt{\frac{2P_{LO}\eta B}{h\nu}} \tag{11}$$

where $e$ is the elementary charge constant, $G$ is the transimpedance gain, $P_{LO}$ is the local oscillator power, $h$ is Planck's constant, $\nu$ is the carrier frequency, $\eta$ is the quantum efficiency, and $B$ is the single-sided bandwidth with a corresponding averaging time $T = (2B)^{-1}$ [3]. The peak signal, $V_{peak}$, is

$$V_{peak} = 2e\eta G\sqrt{\frac{P_{LO}P_s}{(h\nu)^2}} \tag{12}$$

where $P_s$ is the total received signal-comb power on the detector[3]. Substitution of (11) and (12) into (10) using $\tau = (2B)^{-1}$ and defining the total number of integrated signal photons in each measurement as $n_s = P_s T/(h\nu)$ gives

$$\sigma_t = \frac{1}{2\ln(2)}\frac{\tau_p}{\sqrt{\eta n_s}}. \tag{13}$$

Then, converting to range, the shot noise contribution to the range deviation is

$$\sigma_R = \left(\frac{c}{2}\right)\frac{1}{\ln(2)}\frac{\tau_p}{2\sqrt{\eta n_s}}. \tag{14}$$

In the actual implementation, a single heterodyne signal is insufficient as we need two measurements to normalize out the fluctuations in the received signal power (and therefore in $V_{peak}$). Therefore, the timing discriminator generates two time-displaced copies of the tracking comb pulses to generate the two displaced discriminator signals $V_1$ and $V_2$, each of which has the same Gaussian shape and noise jitter as given above although with half the total power. The timing discriminator signal is given by the linear combination, $S = (V_1 - V_2)/(V_1 + V_2)$. An identical analysis for the shot-noise limited timing jitter and ranging jitter yields exactly the same equations (13) and (14), where $n_s$ is the total number of signal photons.

The quantity in Eq. (14) is plotted in Figure 3c as a function of $P_S$ at a set averaging time of $T = 200$ ms for the quantum efficiency $\eta \approx 0.79$ and $\tau_p = 355$ fs, which is the time-bandwidth limited pulse duration for the measured comb pulse spectral FWHM of 10.1 nm.

The above derivation assumed the optimal situation that the two comb pulses have no differential chirp, are Gaussian, have unity mixing efficiency and the additive detector noise is zero. In Eq. (1), we include the numerical factor $C$ to account for these and any other effects that increase the noise above this quantum-limited floor. In our case, the FWHM of the timing discriminator signals for pulses with zero differential chirp should be $2\tau_p = 709$ fs, based on the measured spectral widths of the comb pulses, but we measure 824 fs for the data of Figure 3c, indicating a differential dispersion between the pulses and corresponding timing jitter penalty of $(824/709)^2 = 1.35$ (assuming the broadening both reduces the slope and the peak height, preserving the area). In addition, there is an SNR degradation of a factor of 1.39 due to detector noise, measured non-idealities in interferogram slopes, and spectral overlap. Finally, we measure an additional additive noise penalty of 1.065 in the RF chain. Combined, these imperfections give an expected value of $C = (1.35)\times(1.39)\times(1.065) = 2.0$, to be compared to the value of $C = 2.16$ measured from the upward displacement of the data points compared to the theoretical curve in Figure 3c. These two values agree to within 8%. For the data of Figure 4, the additional round-trip length of fiber optics in the path to the rail-mounted retroreflector added chirp to the signal comb that led to an additional 1.49× penalty.

For the conventional dual-comb ranging using linear optical sampling, a similar analysis has been provided in Appendix B of Ref. 1. The performance curves in Figure 5 are based on these equations and assume no additional penalties.

## REFERENCES (METHODS)

**Extended Data Table 1:** Comparison of parameters for FCMW ranging, conventional dual-comb ranging and the current TPFC-based dual-comb ranging. Scaling relationships and values from the literature are used to illustrate the trade-offs. All systems have a maximum range set by the laser coherence length.

$B$: 3-dB bandwidth, C: chirp rate, $c$: speed of light (for simplicity ignoring the group index of air), $l_{coh}$: laser coherence length, $f_{rep}$: repetition frequency of comb, $\Delta f_{rep}$: offset in comb repetition frequencies, $S$: pulse scan rate.

| Technique | Resolution, $\Delta R$ | | Non-Ambiguity Range, $R_{NA}$ | | Power penalty, $P_P$ | | Max. Update Rate, $T^{-1}$ | |
|---|---|---|---|---|---|---|---|---|
| **FMCW Ranging**[a] | | | | | | | | |
| **System 1**[b,c] $B = 5$ THz (40 nm at 1550 nm) $C = 5$ THz/s (40 nm/s) | $\dfrac{c}{2B}$ | 30 µm | $\sim \dfrac{cC}{4B}$ | $\gg l_{coh}$ | ~1 | 0 dB | $\dfrac{C}{B}$ | 1 Hz |
| **System 2**[b,d] $B = 1.3$ THz (10 nm 1535 nm) $C = 12.5$ THz/s (100 nm/s) | | 115 µm | | | | | | 9.6 Hz |
| **Conventional Dual-Comb Ranging** | | | | | | | | |
| **Microcomb**[e]: $B = \sim 2.5$ THz $f_{rep} = 100$ GHz $\Delta f_{rep} = 100$ MHz | | 60 µm | | 0.15 cm | | 14 dB | | ~1 GHz |
| **Er:fiber comb**[f]: $B = 1.2$ THz (10 nm at 1560 nm) $f_{rep} = 200$ MHz $\Delta f_{rep} = 2$ kHz | $\dfrac{c}{2B}$ | 125 µm | $\dfrac{c}{2f_{rep}}$ | 75 cm | $\sim \dfrac{f_{rep}}{B}$ | 38 dB | $\dfrac{f_{rep}^2}{4B}$ For margin, in practice this is often[h] $\dfrac{f_r^2}{16B}$ | 8.3 kHz |
| **Ti:Sapph comb**[g]: $B = 0.3$ THz (0.6 nm at 785 nm) $f_{rep} = 513$ MHz $\Delta f_{rep} = 130$ kHz | | 510 µm | | 29 cm | | 27 dB | | 220 kHz |
| **Dual-Comb Ranging with the TPFC** $B = 1.2$ THz (10 nm at 1560 nm) Acquisition scans over $0 < \Delta X < f_{rep}^{-1}$ Update rate assumes $\Delta X = 1$ ns (15 cm) and $\Delta \dot{X} = 40$ n/s | $\dfrac{c}{2B}$ | 125 µm | $\dfrac{c}{2f_{rep}}$ | 75 cm | Acquisition: $\Delta X B$  Tracking: 1 | 0 to 30 dB  0 dB | Acquisition: $S \Delta X$  Tracking: $\dfrac{f_{rep}}{4}$ | 40 Hz  50 MHz |

[a] Scaling for FMCW based on Barber, Z. W., Babbitt, Wm. R., Kaylor, B., Reibel, R. R. & Roos,. Appl Opt 49, 213–219 (2010).
[b] The use of tradenames in this manuscript is necessary to specify experimental results and does not imply endorsement by the National Institute of Standards and Technology.
[c] System 1 specifications from Bridger Photonics datasheets (https://www.bridgerphotonics.com/).
[d] System 2 specifications from Luna Optical Backscatter Reflectometer (OBR) 4600 datasheet (https://lunainc.com).
[e] Values taken from Trocha, P. et al.. Science 359, 887–891 (2018). FWHM bandwidth estimated to be at 2.5 THz from the manuscript figures.
[f] Values taken from Er:fiber comb system used also for the dual-comb system presented here.
[g] Values taken from Mitchell, T., Sun, J., Sun, J. & Reid, D. T.. Opt. Express 29, 42119–42126 (2021).
[h] This expression often appears with only a factor of two in the denominator. However, if the widths are all FWHM and assuming a filter that is not infinitely sharp, there would be significant aliasing in the signal if operated under these conditions leading to systematic range errors. We have used a factor of "4", which is still overly optimistic and the factor of "16" in the denominator is a more practical choice.

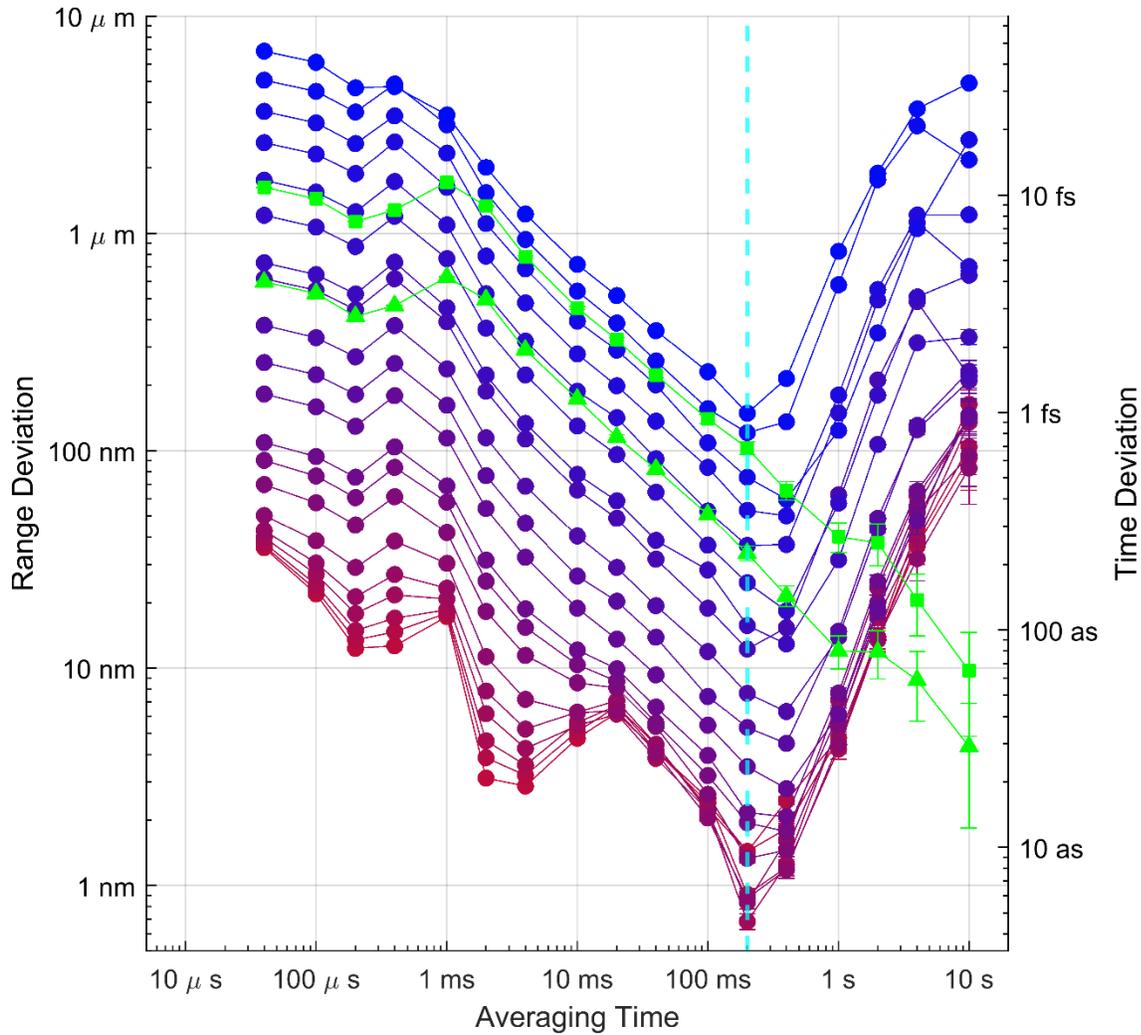

**Extended Data Figure 1:** Range deviation (left axis) and time deviation (right axis) of dual comb range measurements from a fixed reflection for signal-comb powers from 980 nW (top curve) down to 0.33 pW (bottom curve) with the following power levels ± 10%: 980 nW, 190 nW, 86 nW, 38 nW, 21 nW, 9.6 nW, 1.8 nW, 1.2 nW, 390 pW, 200 pW, 89 pW, 33 pW, 23 pW, 8.5 pW, 4.1 pW, 1.9 pW, 990 fW, 550 fW and 330 fW. The vertical dashed cyan line indicates the 200-ms averaging time for the data in Fig. 3c. Beyond 200-ms, the range deviation increases due to temperature-induced fluctuations in the fiber path up to the fixed reflection. In addition, the range and time deviations for the difference between the absolute range from the tracking comb timing and the relative range from the unwrapped carrier phase shift, as given in Figure 4. The data are from the time periods of 60 to 100 seconds at 3.2 pW (green squares) and the period from 110 seconds to 150 seconds at 32 pW (green triangles), respectively. For these data, the differential chirp between the signal and TPFC pulses was larger, leading to an additional 1.5x penalty in $C$ and thus lie slightly above the curves at the same power for ranging off the fixed reflection (solid circles). However, because the path length variation is common mode, the difference continues to average down beyond 200 ms.

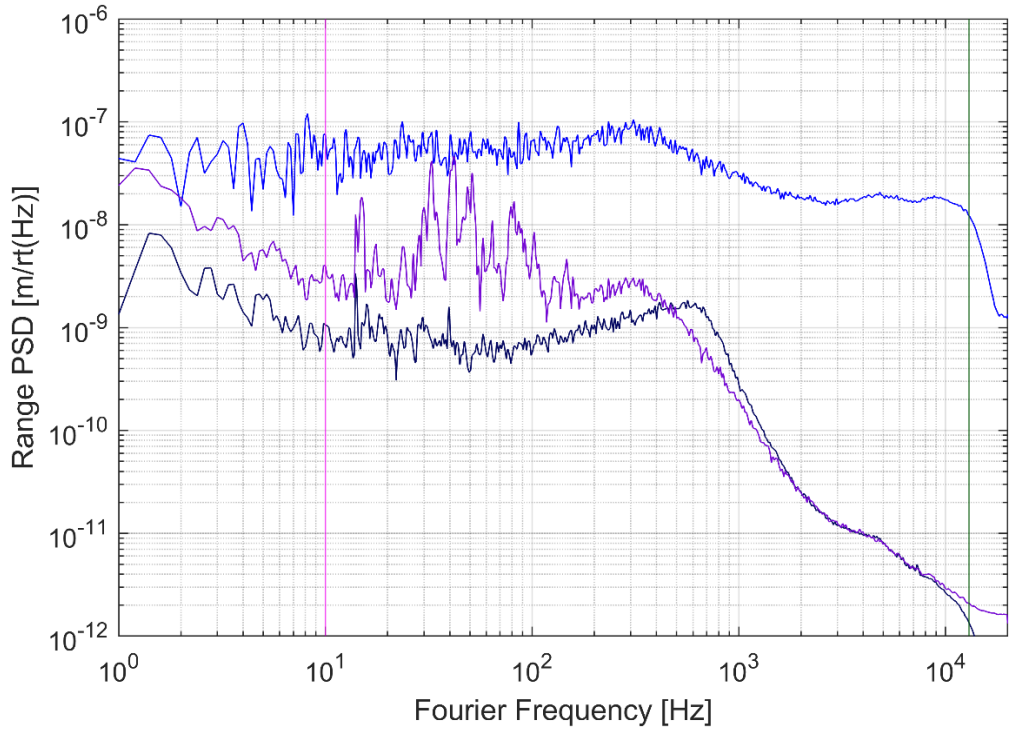

**Extended Data Figure 2:** Range Power Spectral Density (PSD) for the data from Figure 4 over the period of 60 s to 100 s at 3.2 pW return power for X(t) from the tracking comb (dark blue trace) and the unwrapped carrier-phase $\theta(t)$ (purple trace). Also shown is the noise floor for the unwrapped carrier phase (black trace). The vibrations of the nominally immobile retroreflector can be clearly seen in the carrier-phase data. At the low average power of 3.2 pW, the tracking dual-comb range shot-noise limited noise floor lies just above the minimal vibrations seen here. The vertical magenta line indicates the maximum 10 Hz update rate of FMCW while the vertical dark green line indicates the 13 kHz cutoff imposed by the 26 kHz measurement rate for the range data.